\documentclass[prb,aps, twocolumn, amsmath,amssymb,superscriptaddress]{revtex4}

\usepackage{float}
\usepackage{amsmath}
\usepackage{amssymb}
\usepackage{amsfonts}
\usepackage{euscript}
\usepackage{enumerate}
\usepackage{hhline}
\usepackage{pslatex}
\usepackage{tabularx}
\usepackage{threeparttable}
\usepackage[usenames,dvipsnames]{xcolor}
\usepackage[colorlinks,linkcolor=red,anchorcolor=blue,citecolor=blue]{hyperref}

\usepackage{graphicx}
\usepackage{dcolumn}
\usepackage{bm}
\usepackage[sort&compress]{natbib}
\usepackage{multirow} 

\newcolumntype{L}[1]{>{\raggedright\arraybackslash}p{#1}}
\newcolumntype{C}[1]{>{\centering\arraybackslash}p{#1}}
\newcolumntype{R}[1]{>{\raggedleft\arraybackslash}p{#1}}

\makeatletter

\newcommand{\ket}[1]{|  #1 \rangle }

\renewcommand{\@biblabel}[1]{#1. }

\renewcommand{\@dotsep}{500}
\renewcommand{\@pnumwidth}{0em}
\renewcommand{\l@figure}[2]{
\@dottedtocline{1}{1.5em}{2em}{Figure #1}{}\vspace{15pt}}

\begin{document}

\title{Telecom band quantum dot technologies for long-distance quantum networks}

\author{Ying Yu}

\affiliation{State Key Laboratory of Optoelectronic Materials and Technologies, School of Physics, School of Electronics and Information Technology, Sun Yat-sen University, Guangzhou 510275, China}

\author{Shunfa Liu}

\affiliation{State Key Laboratory of Optoelectronic Materials and Technologies, School of Physics, School of Electronics and Information Technology, Sun Yat-sen University, Guangzhou 510275, China}

\author{Chang-Min Lee}
\affiliation{Department of Electrical and Computer Engineering and Institute for Research in Electronics and Applied Physics, University of Maryland, College Park, Maryland 20742, USA}
\affiliation{Joint Quantum Institute, NIST/University of Maryland, College Park, MD 20742, USA}

\author{Peter Michler}
\affiliation{Institut f$\ddot{u}$r Halbleiteroptik und Funktionelle Grenzfl$\ddot{a}$chen (IHFG), Center for Integrated Quantum Science and Technology (IQST) and SCoPE, University of Stuttgart, Allmandring 3, 70569 Stuttgart, Germany}

\author{\\Stephan Reitzenstein}
\affiliation{Institute of Solid State Physics, Technische Universitl$\ddot{a}$t Berlin, 10623 Berlin, Germany}

\author{Kartik Srinivasan}
\affiliation{Microsystems and Nanotechnology Division, Physical Measurement Laboratory, National Institute of Standards and Technology, Gaithersburg, MD 20899, USA}
\affiliation{Joint Quantum Institute, NIST/University of Maryland, College Park, MD 20742, USA}

\author{Edo Waks}
\affiliation{Department of Electrical and Computer Engineering and Institute for Research in Electronics and Applied Physics, University of Maryland, College Park, Maryland 20742, USA}
\affiliation{Joint Quantum Institute, NIST/University of Maryland, College Park, MD 20742, USA}

\author{Jin Liu}
\thanks{liujin23@mail.sysu.edu.cn}
\affiliation{State Key Laboratory of Optoelectronic Materials and Technologies, School of Physics, School of Electronics and Information Technology, Sun Yat-sen University, Guangzhou 510275, China}

\date{\today}

\begin{abstract}
\noindent A future quantum internet is expected to generate, distribute, store and process quantum bits (qubits) over the globe by linking different quantum nodes via quantum states of light. To facilitate the long-haul operations, quantum repeaters, the building blocks for a long-distance quantum network, have to be operated in the telecom wavelengths to take advantage of both the low-loss fiber network and the well-established technologies for optical communications. Semiconductors quantum dots (QDs) so far have exhibited exceptional performances as key elements, i.e., quantum light sources and spin-photon interfaces, for quantum repeaters, but only in the near-infrared (NIR) regime. Therefore, the development of high-performance telecom-band QD devices is highly desirable for a future solid-state quantum internet based on fiber networks. In this review, we present the physics and the technological developments towards epitaxial QD devices emitting at the telecom O- and C-bands for quantum networks by using advanced epitaxial growth for direct telecom emission, and quantum frequency conversion (QFC) for telecom-band down-conversion. We also discuss the challenges and opportunities in the future to realize telecom QD devices with improved performances and expanded functionalities by taking advantage of hybrid integrations.
\end{abstract}

\maketitle

\section{Introduction}
\begin{center}
	\begin{figure*}
		\includegraphics[width=0.6\linewidth]{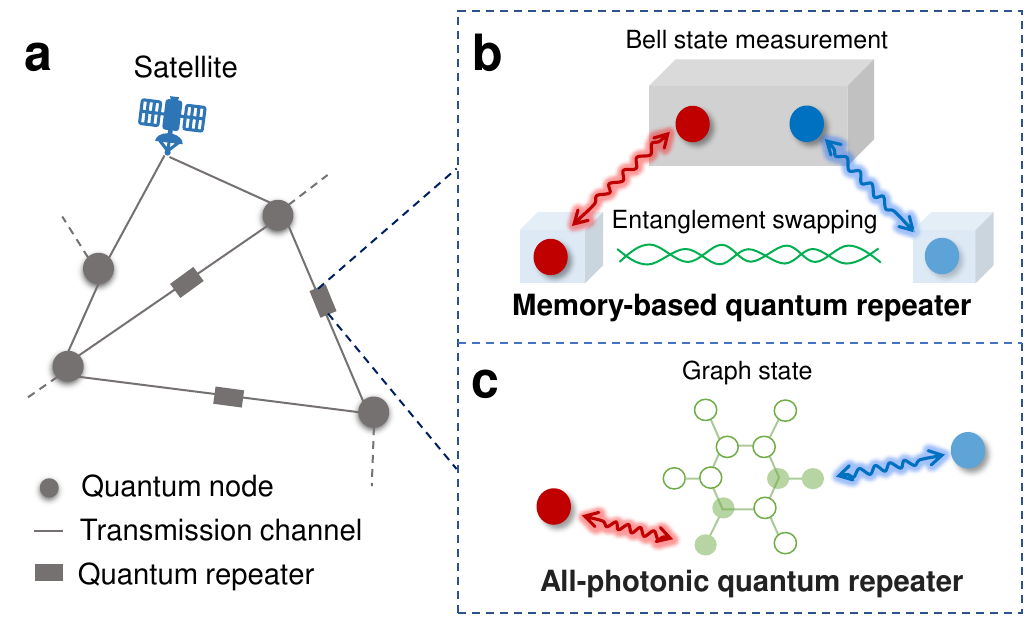}
		\caption{\textbf{Global quantum network connected by quantum states of light.} (a) Illustration of a quantum network consisting of quantum nodes, quantum repeaters, as well as quantum/classical transmission channels; (b) Sketch of a memory-based quantum repeater, which can store and retrieve quantum information by independently creating entanglement for individual quantum nodes and then store this entanglement in quantum memory--intermediaries between different quantum systems; (c) Sketch of all-photonic quantum repeater, which introduces a repeater graph state that consists of a complete connected subgraph of N core photons (N = 6 in our example) to implement entanglement swapping in the secondary nodes. }
		\label{fig:Fig1}
	\end{figure*}
\end{center}

Emerging quantum networks\cite{kimble2008quantum,simon2017towards} provide unprecedented opportunities and challenges for exploring fundamental quantum physics and developing quantum technologies across a variety of the research and application frontiers such as quantum communication, quantum computation and quantum metrology. A photonic quantum network\cite{simon2017towards,frohlich2013quantum}, as shown in Fig.~\ref{fig:Fig1}a, consists of a global distribution of quantum information processors, quantum repeaters and quantum/classical transmission channels, in which stationary qubits are generated and processed in quantum information processors, while flying qubits (quantum states of light) transmit the quantum information through transmission channels such as the air, a vacuum of space, fiber networks or on-chip quantum circuits. The adjacent nodes and remote nodes in the quantum network are connected by quantum relays or quantum repeaters. Quantum relays are typically operated in a quantum teleportation configuration to expand the quantum communication distance between different nodes without involving quantum memories, as required for a fully fledged quantum repeater network \cite{PhysRevLett.92.047904}. To reach an arbitrarily long distance quantum network, quantum repeaters\cite{briegel1998quantum,arXiv:221210820 2022}, as shown in Fig.~\ref{fig:Fig1}b, are proposed by either employing entangled photon pairs with quantum memories or multi-photon entangled cluster graph states. In memory based quantum repeaters, a long-distance  quantum channel is divided into shorter segments. Each segment is connected through entanglement swapping\cite{zukowski1993event} via Bell-state measurements (BSMs). Entangled quantum states are stored and purified in memories\cite{chen2008memory} before entanglement swapping connects all these segments. To implement a quantum memory\cite{lvovsky2009optical,lpor.202100219}, a robust light-matter interface between the flying photonic states and the long-lived matter states, long storage times and multiplexed operation are required. The underlying process is often based on atomic lambda ($\Lambda$) systems (atom assembles\cite{hammerer2010quantum}, nitrogen-vacancy center in diamonds\cite{dutt2007quantum}, rare-earth ion doped solids\cite{yang2018multiplexed}, etc.), in which the photon-spin entanglement is mapped into photon-photon entanglement.The aforementioned spin memories have been verified with high storage efficiency and long storage time\cite{lpor.202100219}, while the orbital memories (often ladder systems) offer fast and noise-free operation\cite{davidson2022fast}. For a recent overview and quantitative assessment of the performance of a memory-based quantum repeater system we refer to P. van Loock et al.\cite{loock2022extending}. In this work, the performances of basic quantum repeater links for different platforms, i.e. quantum dots (QDs), trapped atoms and ions, and color centers in diamond, are evaluated and compared for state-of-the-art experimental parameters as well as for parameters that can in principle reached in the future. It is shown that the “repeaterless” bound in secret key rate can be exceeded for link coupling efficiencies of 60\%, clock rates
of 1 GHz and spin memory coherence times of 0.3 ms.

Another approach is based on all-photonic quantum repeaters\cite{azuma2015all} which use logically-encoded multi-photon states such as large repeater graph states, as shown in Fig.~\ref{fig:Fig1}c. All-photonic quantum repeaters can send and receive the heralding signals for the entanglement swapping in the same repeater node. The quantum nodes are linked via non-classical states of photons emitted by quantum light sources, providing redundancy against photon loss and the probabilistic nature of photonic BSMs.

To obtain long-distance quantum networks\cite{simon2017towards}, different transmission channels depending on the required propagation distance are considered, such as optical fiber links (\textless 500 km), fiber-based quantum repeaters (500-2000 km), satellite links (\textgreater 2000 km), or some combination of these approaches. For high transmission rates, it is desirable that the quantum light sources feature very high brightness and operate at telecommunication wavelengths, i.e., within the telecom O-band ($\approx$1310 nm) or C-band ($\approx$1550 nm), due to the extremely-low transmission loss (0.35 dB/km at 1310 nm and 0.2 dB/km at 1550 nm) in the full-blown optical fiber links\cite{michler2017quantum} as well as the relatively low-background solar irradiance and Rayleigh scattering in the free-space and satellite-based quantum communication\cite{liao2017long,Avesani2021full}. Furthermore, manipulations of quantum states of light in the telecom bands can greatly benefit from the well-established technologies developed for classical optical communications and silicon photonics, e.g., high-performance semiconductor laser sources\cite{matsui2021low}, wide-bandwidth optical modulators\cite{he2019high} and high-speed photon detectors\cite{zhang2015advances} are readily available at telecom bands. For example, the single-photon detection efficiency as high as 93\% at 1550 nm is now available through the development of superconducting nanowire single-photon detectors (SNSPDs)\cite{Marsili2013detector}.

Semiconductor QDs exhibit exceptional performances as both quantum light sources and spin-photon interfaces for developing quantum repeaters (See Box.1). The universally used spontaneous parametric down-conversion (SPDC) and attenuated laser source in quantum communications inevitably suffer from the compromise between the source brightness and single-photon purity, which severely limits the device performance in terms of transmission distance and communication speed. In contrast, semiconductor QDs are able to generate single-photon Fock states in a deterministic manner. The state-of-the-art QD single-photon sources in the near-infrared (NIR) regime exhibit simultaneous high-degrees of end-to-end source brightness (with a record number of 57\%), single-photon purity and photon indistinguishability, which already outperform SPDC sources\cite{tomm2021bright}. The benefits of exploiting QD sources as single-photon and entangled-photon pairs for quantum communications are currently under very active investigations\cite{AQT2022QDreview, npj2022QDreview}. Very recent modeling suggests that a telecom band QD single-photon source featuring a collection efficiency over 30$\%$ could outperform the commercial decoy-states scheme, and the implementations of twin-field QKD protocol can further improve the QKD distance \cite{npj2022QDreview}. In addition, the availability of electrical-injection of the QD source is highly beneficial for applications due to the avoidance of bulky and expensive laser system and much higher clock rates beyond 1 GHz\cite{PhysRevApplied.13.054052}. Towards a memory-based quantum repeater, the state-of-the-art NIR QD sources with a pair collection probability of up to 0.65(4) is one order of magnitude brighter than the SPDC source (typically with a pair rate of \textless 0.1)\cite{liu2019solid}. Potentially working as a quantum memory, spin states of confined electrons/holes in QDs can be optically initialized, manipulated and read out with high-fidelity and fast-speed, which enables the realization of spin-photon entanglement\cite{gao2012observation} in single QDs, and even spin-spin entanglement between remote QDs\cite{delteil2016generation}. In the context of all-photonic quantum repeaters, a record-performance for the deterministic generation of photonic cluster state with an entanglement length up to 10-photons and a generation rate of 0.5 GHz was recently demonstrated in an InGaAs QD emitting at 950 nm\cite{Cogan2023cluster}. Realistic optimizations of the sources could push the entanglement length as long as 50-photons with a generation rate as high as 2 GHz\cite{Cogan2023cluster}. Such performances could immediately be employed to generate high-dimensional cluster states for the long-sought-after demonstrations of all-photonic quantum repeaters and distributed quantum computation. However, all the aforementioned QD devices were operated at NIR regime, experiencing an absorption at least one order of magnitude stronger than that in C-band. Thus, it is highly desirable yet challenging to transfer such exceptional performances from the NIR regime to the telecom bands for building long-distance quantum internet based on fiber networks. Here we discuss the past and on-going research activities towards semiconductor-based quantum networks and quantum repeaters in the telecom bands by exploring both advanced epitaxial growth and quantum frequency conversion (QFC). We also summarize the state-of-the-art and identify the opportunities toward real quantum networks by exploring the potential of hybrid integrated quantum photonics.

\ \hspace*{\fill} \

\textcolor{violet}{\textbf{Box 1: Epitaxial QDs for quantum light sources and spin-photon interfaces}}

\begin{center}
	\begin{figure}
		\includegraphics[width=\linewidth]{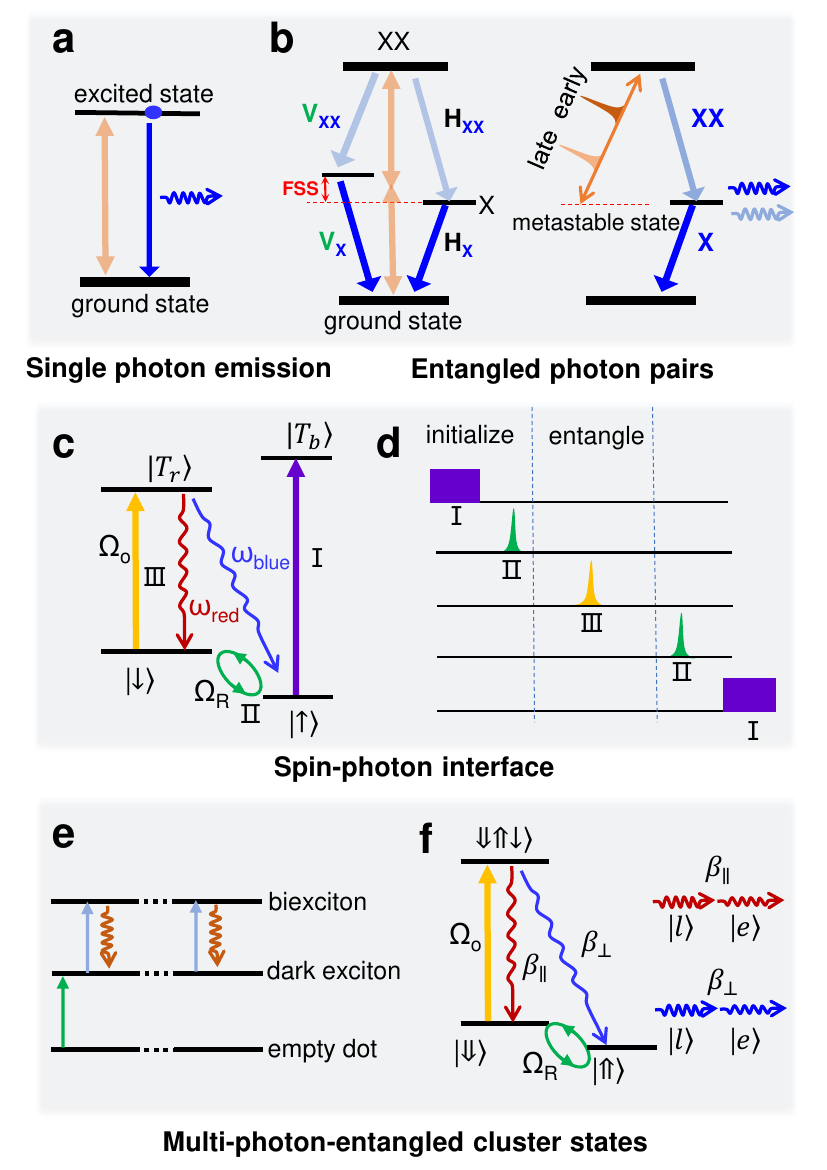}
		\caption{\textcolor{violet}{\textbf{Epitaxial quantum dots (QDs) for building quantum nodes. }(a) Illustration of a single photon generated via spontaneous radiation between the discrete energy levels in a single QD; (b) Illustration of polarization (left) or time-bin (right) entangled photon pairs generated via the biexciton (XX)-exciton (X) cascaded radiative processes in a single QD; (c) Photon-spin interface in a single QD with a two-$\Lambda$ energy-level diagram and the allowed optical transitions in Voigt geometry under an external magnetic field ($\rm B_x$) applied perpendicular to the growth direction; (d) Timing diagram of the pulse sequence for the spin-photon interface in QDs, including spin initialization (\uppercase\expandafter{\romannumeral1}), entanglement generation (\uppercase\expandafter{\romannumeral2}), and spin read-out (\uppercase\expandafter{\romannumeral3}); (e-f) Multi-photon-entangled cluster states created in a single QD by using a dark exciton\cite{schwartz2016deterministic} (e) or a hole spin\cite{tiurev2022high} (f). In (e) the cluster state is based on polarization entanglement and in (f) on time-bin entanglement. The wavy arrows in (a-f) represent photons. Images in (e-f) are adapted with permission from Refs. [\cite{tiurev2022high,schwartz2016deterministic}], respectively.}}
		\label{fig:Fig2}
	\end{figure}
\end{center}

\textcolor{violet}{Semiconductor QDs, regarded as artificial atoms due to their discrete electronic energy levels, offer a promising way to create quantum light sources and generate light-matter entanglement in a single physical system. Single-photon generation here relies on spontaneous transitions between the discrete energy levels, with the production of a photon at each decay from an excited state (Fig.~\ref{fig:Fig2}a). Polarization or time-bin entangled photon pairs are generated via the biexciton (XX)-exciton (X) cascade radiative processes in a single QD, as shown in Fig.~\ref{fig:Fig2}b. }

\textcolor{violet}{QDs feature a particularly high oscillator strength for the spin-photon interface, providing further opportunities for spin-photon entanglement generation\cite{gao2012observation}. A single spin can be realized using a dark exciton\cite{schwartz2015deterministic} or a single charge (either a hole\cite{gerardot2008optical} or an electron\cite{berezovsky2008picosecond}) injected into the QD by delta-doping or electrically gated-devices and can be prepared/readout via laser pulses. The spin-photon interference arises in the optical spectroscopy of a $\Lambda$ system consisting of three states, which is formed in Voigt geometry under an external magnetic field ($\rm B_x$) applied perpendicular to the growth direction\cite{gerardot2008optical,berezovsky2008picosecond}, as illustrated in Fig.~\ref{fig:Fig2}c. With the excitation of the ground state of the QD ($\ket{\downarrow}$ denotes spin antiparallel to the magnetic field) into the trion state ($\ket{\rm{Tr}}$), the photon states of two different radiative recombination paths ($\rm{\omega_{red}}$ and $\rm{\omega_{blue}}$) can be naturally entangled with the matter spin states ($\ket{\downarrow}$ and $\ket{\uparrow}$), creating spin-photon entangled states. The typical measurement sequence including spin initialization (\uppercase\expandafter{\romannumeral1}), entanglement (\uppercase\expandafter{\romannumeral2}), and read-out (\uppercase\expandafter{\romannumeral3}) is illustrated in Fig.~\ref{fig:Fig2}d. The optical spin rotation and echo techniques (green circle) are usually explored to prolong the spin coherence time.}

\textcolor{violet}{Linear cluster states can be generated by repeated and timed optical excitations of a confined spin in a single semiconductor QD, such as dark exciton or hole spin, as shown in Figs.~\ref{fig:Fig2}e-f. The precessing spin will act as an “entangler” and entangles the consequently emitted photons, leading to the explicit demonstration of multi-photon-entangled cluster states\cite{tiurev2022high,schwartz2016deterministic}. Both degrees of freedom, time-bin (Fig.~\ref{fig:Fig2}e) and polarization (Fig.~\ref{fig:Fig2}f) have been used for first demonstrations of photonic cluster states. More recently, by deterministic generation a heavy-hole spin in a QD, it was shown that entanglement can persist over ten photons, with the indistinguishability that is still keeping above 90\%\cite{Cogan2023cluster}. Spin-photon entanglement has been also injected into micropillars to generate linear-cluster states with much higher rates\cite{coste2023high}}.

Furthermore, QD molecule devices\cite{schall2021bright} can also be interesting for qubit storage\cite{de2011exciton} and cluster-state generation\cite{economou2010optically}. A single driven electron in a QD also features a collective mode of a nuclear spin ensemble that is proposed as a large nuclear spin register for quantum memory\cite{gangloff2019quantum}. Moreover, the storage of deterministic single photons emitted from a QD (879.7 nm) has been demonstrated in a polarization-maintaining solid-state quantum memory based on $\rm Nd^{3+}:YVO_{4}$ crystals\cite{Tang2015storage}.

\section{Growth methods of QDs emitting at telecom wavelengths}

The most common approach of growing QDs is based on the Stranski-Krastanov (S-K) mode using molecular beam epitaxy (MBE) or metal-organic vapor phase epitaxy (MOVPE). Two material systems are exploited to shift the emission wavelength from the NIR to the telecom O- and C-bands: In(Ga)As/GaAs and InAs/InP. However, in general the self-assembled S-K growth method results in randomly positioned QDs with a significant inhomogeneous broadening of the QD emission range. This inhomogeneity, together with the random positions, strongly impair the production yield of devices based on single QDs, imposing a formidable challenge to the scalability. Thus, in addition to S-K growth mode, we will also address droplet epitaxy, which significantly reduces the QD density, the lattice strain and the exciton fine structure splitting (FSS) for polarization-entangled photon pair generation, as well as site-controlled growth, which overcomes the spatial randomness of S-K growth mode.

\subsection{Stranski-Krastanov growth}

\begin{center}
	\begin{figure*}
		\includegraphics[width=0.95\linewidth]{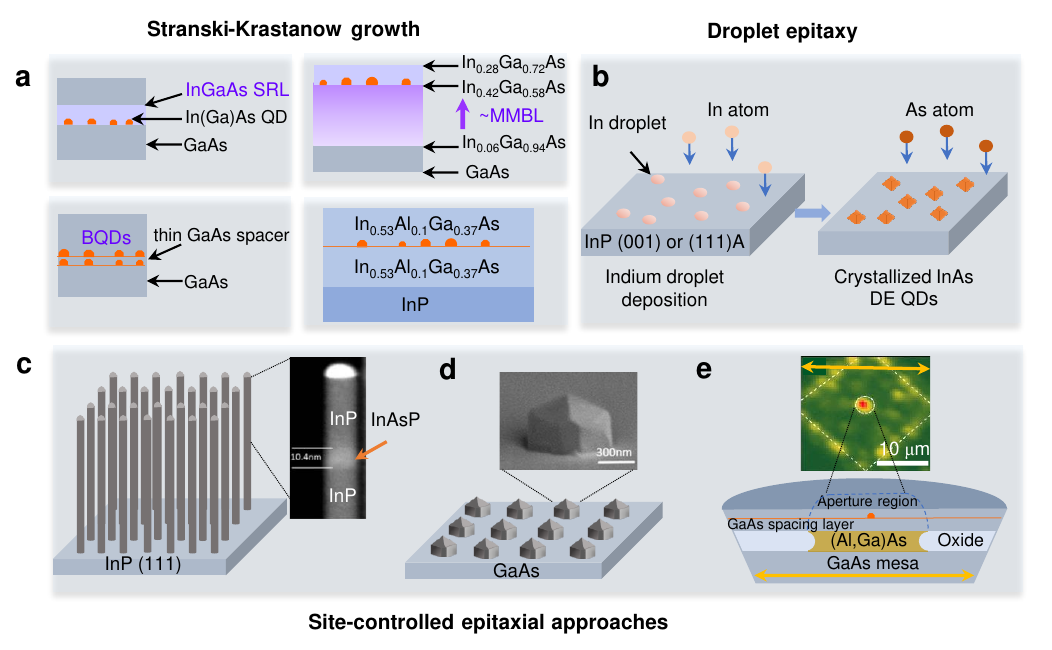}
		\caption{\textbf{Epitaxial growth for telecom QDs.} (a) Sketches of Stranski-Krastanow growth of QDs emitted at telecom wavelengths, including (1) InGaAs QDs with InGaAs SRL (upper left), (2) InAs QDs above a metamorphic InGaAs matrix (upper right), (3) bilayer QDs (bottom left) and (4) InAs/InAlGaAs/InP QDs (bottom right); (b) Schematic of growth and fabrication stages of droplet etching (DE)-based QDs. (c) Schematic of InP nanowires arrays grown on InP (111)B. Each photonic nanowire waveguide contains a single InAsP/InP QD, together with a high angle annular dark field (HAADF) scanning TEM image of an  $\rm InAs_{0.5}P_{0.5}$ QD in InP nanowire\cite{haffouz2018bright}. (d) Schematic of a mesa-top single QD (MTSQD) array. A magnified SEM of a single mesa bearing MTSQD is shown in the upper SEM image\cite{zhang2020planarized}; (e) Schematic of the buried-stressor structure\cite{strittmatter2012lateral}, together with the micro-PL mapping results after the site-controlled growth\cite{grosse2020development}; Reprinted with permission from Refs. [\cite{haffouz2018bright,zhang2020planarized,strittmatter2012lateral,grosse2020development}]}
		\label{fig:Fig3}
	\end{figure*}
\end{center}

\textbf{In(Ga)As/GaAs QDs}. The lattice mismatch between InAs and GaAs is 7.2\%, inducing a typical emission wavelength of $\approx$920 nm at 4K. To redshift the ground-state transition energy of In(Ga)As QD from the NIR to the telecom spectral range, the primary approach is to decrease the lattice mismatch for increasing the QD size. One concept is to apply a strain-reduce layer (SRL) on the top of In(Ga)As QDs during growth to minimize strain and surface energy\cite{tatebayashi2001over}. An $\rm In_{x}Ga_{1-x}As$ (0.1 $\textless$ x $\textless$ 0.3) layer, shown in upper left of Fig.~\ref{fig:Fig3}a, is used to cover the QDs instead of capping with GaAs after the formation of QDs\cite{kettler2016single}.
Single-photon emission with a high multi-photon suppression of $g^{(2)}(0)$=0.027 $\pm$ 0.005, and a post-selected visibility of 96$\%$ (raw visibility: 12$\%$) was reported, using InAs QD with SRL (telecom O-band) combined with p shell excitation\cite{srocka2020deterministically}. Another option to redshift to telecom wavelength is capping InAs QDs with a thin $\rm GaAs_{1-x}Sb_{x}$ layer\cite{Orchard2017silicon}. However, the QD linewidths are $\approx$3-5 meV, which attributed to more fluctuating carrier occupancy of states in InAs/$\rm GaAs_{1-x}Sb_{x}$ system with type-\uppercase\expandafter{\romannumeral2} band energy, leading to low-coherence photons.

Another approach is to increase the lattice constant of the buffer layer below the QDs, as presented in upper right of Fig.~\ref{fig:Fig3}a. In this case, the InAs QDs are deposited on an InGaAs metamorphic buffer layer (MMBL)\cite{portalupi2019inas}. The lattice mismatch between $\rm In_{x}Ga_{1-x}As$ and InAs is reduced by increasing the indium content of MMBL, resulting in larger dot size and lower band gap of the QD material which allows to reach emission in the telecom C-band\cite{paul2017single}. Alternatively, strain-coupled bilayer QDs (BQDs), shown in bottom left of Fig.~\ref{fig:Fig3}a, can be introduced\cite{liu2015electronic}. As its name suggests, two vertically coupled QD layers separated by a thin GaAs or $\rm Al_{x}Ga_{1-x}As$ spacer layer (with a thickness of less than 10 nm) are implemented to increase the actual size of the QDs and therefore extend the wavelength to the telecom O- and C-bands. Combined with a resonant pumping scheme, an on-demand single-photon source\cite{nawrath2019coherence} and entangled photon pair source\cite{zeuner2021demand} based InAs/GaAs QDs with MMBLs emitting in the telecom C-band, have been demonstrated.

\textbf{InAs/InP QDs}. Compared with lattice mismatch (7.2\%) in InAs/GaAs QDs, the lattice mismatch between InAs and InP is much smaller (3.2\%). Thus, this type of InAs QDs is an interesting candidate to emit single photons at the telecom C-band without any strain engineering layer. However, the smaller lattice mismatch also results in high QD density (often larger than $\rm10^{10}\ cm^{-2}$)\cite{marchand1997metalorganic} and relatively large QDs (with emission beyond 1600 nm at room temperature\cite{fafard1996inas}), as well as As/P exchange at the InAs/InP interface\cite{sobiesierski1997p}. A growth method based on ripening of InAs quantum sticks (QSs), which is triggered by the sample cooling under arsenic overpressure, has been developed to reduce the QD density\cite{dupuy2006low}. Additionally, an ultrathin interlayer ($\rm In_{x}Ga_{1-x}As$ or InAlGaAs) underneath the InAs layer\cite{anantathanasarn2005wavelength} or double-InP-capping method\cite{takemoto2004observation} can be introduced to achieve the C-band emission, as illustrated in bottom right of Fig.~\ref{fig:Fig3}a. The latter approach can also reduce QD density and suppress the As/P exchange at the InAs/InGaAsP interface.
\subsection{Droplet epitaxy}

Due to the high density of InAs/InP QDs in S-K growth mode of typically larger than $\rm10^9\ cm^{-2}$, sub-$\rm \mu m$ photonic structures are often employed to realize single-QD devices\cite{takemoto2004observation}. In this context, droplet epitaxy (DE) is a very promising alternative to the conventional S-K mode growth\cite{koguchi1991new}. It starts by depositing group $\rm \uppercase\expandafter{\romannumeral3}$ atoms at controlled temperature and flux to form droplets with appropriate density and size on the substrate, followed by a group $\rm \uppercase\expandafter{\romannumeral5}$ deposition to crystallize the initial droplets into $\rm \uppercase\expandafter{\romannumeral3}$-$\rm \uppercase\expandafter{\romannumeral5}$ islands directly (Fig.~\ref{fig:Fig3}b). DE QDs have higher structural symmetry and, thus, lower FSS if compared with S-K QDs, which facilitate the generation of polarization-entangled photon pairs\cite{skiba2017universal}. Single photons\cite{ha2020single} and entangled photon pairs\cite{muller2018quantum} in the C-band have been reported using InAs DE grown on InP by MOVPE and MBE. It is further noted here that also InAs DE QDs grown on InP (111)A, a substrate which has $\rm C_{3v}$ crystal symmetry, have the advantage of high symmetry and low FSS\cite{liu2014vanishing}. Entangled photon pair emission by InAs/InP DE QD with a fidelity of (87 $\pm$ 4)\% was reported with emission at 1550 nm\cite{muller2018quantum}, highlighting a viable path to the QDs-based entangled light sources at telecom wavelength for long-distance quantum network applications.

More interestingly, the so-called “nanodrill” effect of group III droplets at high growth temperatures can be used to create arrays of local droplet etching (LDE) nanoholes that serve as growth templates for QDs. In constrast to the S-K concept, this method does not rely on lattice mismatch between the matrix and QD materials and thus allows for the growth of the strain-free QDs, such as highly symmetric GaAs/AlGaAs QDs. The reported GaAs QDs (emitting at $\approx$780 nm) that formed in the defect and impurity-free AlGaAs nanoholes, yield almost diminishing FSS leading to excellent entangled photon pair generation with   an entanglement fidelity of 0.88(2), and a high single photon indistinguishability of 0.903(3)\cite{liu2019solid}. By embedding GaAs QDs into an n-i-p-diode, single-photon emission with close-to lifetime-limited linewidths and no-blinking, as well as an electron spin initialization with high fidelity and a spin-relaxation time as large as  50 $\mu$s are achieved \cite{zhai2020low}.  For high performance LDE QD emitting at the telecom wavelength, GaSb/AlGaSb\cite{chellu2021highlyuniform} or In(Ga)As/InAlAs/InP\cite{cao2022local} materials will have the opportunity to be established and engineered to the growth of lattice-matched and defect-free QDs.

\subsection{Site-controlled growth}

Due to the self-assembled mechanism, S-K and DE growth modes result in randomly positioned and highly inhomogeneous QD populations. This makes the fabrication of single QD devices with a high scalability very challenging and has led to the development of advanced deterministic QD-device fabrication technologies\cite{liu2021nanoscale}. Overcoming spatial and spectral inhomogeneities already at the growth stages would constitute a major breakthrough, and new site-controlled growth methods may provide a pathway to realize the atomic-scale control for ultimate spectral engineering of QDs. A QD in a nanowire (NW) structure\cite{singh2009nanowire}, promises a high position accuracy because these emitters can grow using a combination of selective-area and vapour-liquid-solid (VLS) epitaxy on patterned [111]-oriented substrates (Fig.~\ref{fig:Fig3}c). The most popular configuration consists of a QD embedded on the NW axis as a disk, forming an axial heterostructure of the type A/B/A, where B shows quantum confinement. Core-shell structures are also implemented to protect the QD from the surface states, which can degrade the optical properties of the NW-QD through nonradiative recombination and spectral diffusion. Notably, single InAsP QDs-in-NWs have exhibited the ability of producing bright single-photon emission with an unprecedented spectral range from 880 nm to 1550 nm\cite{haffouz2018bright}.

Other site-controlled epitaxial approaches, which have been successfully applied to (In)GaAs/GaAs QDs emitting at $\rm \lambda$ \textless 1000 nm, may provide further opportunities for pursuing QDs emitting at telecom wavelengths. The first promising technique is spatially selective growth using substrate-encoded size-reducing epitaxy (SESRE) on spatially regular arrays of patterned nanomesas, as shown in Fig.~\ref{fig:Fig3}d, which enabling single QD arrays with spectral uniformity of $\approx$5 nm\cite{zhang2020planarized}. Another improved site-controlled technology is based on the buried-stressor growth\cite{strittmatter2012lateral}. Here, local strain-engineering by an oxide-aperture leads to the localized formation of single QDs, as illustrated in Fig.~\ref{fig:Fig3}e. Tight site-control, very good optical properties and even resonance fluorescence have been observed for buried-stressor InAs/GaAs QDs emitting at 920-950 nm\cite{grosse2020development}, and it will be interesting to further develop this technique for realizing site-controlled QDs in the O-band by combining it for instance with the SRL concept.

\section{Coupling to telecom band single-mode fibers}

The underlying philosophy in developing telecom band QDs is to take advantage of the mature optical fiber networks to expand the transmission distance of quantum information. For example, the 3 dB propagation distance in single mode optical fiber at 1550 nm is 15 km, while in the 900 nm band it is ten times smaller. Therefore, the single-photons and entangled photon pairs emitted by the telecom-band QDs have to be engineered to force its non-directional spontaneous emission into one specific mode that can then be coupled into a fiber. This can be achieved using microcavities, which enhance the coupling between the QD dipole and the electromagnetic mode confined in the cavity, or with waveguides, which inhibit the emission outside of the waveguide mode. For efficient coupling to fiber, one can first extract the photons to free-space and subsequently couple the far-field emission into single-mode fibers (most natural for geometries that typically emit into free-space, such as micropillars or circular Bragg gratings (CBGs) or instead directly couple the photons to single-mode fibers via near-field coupling (most natural for QDs that emit into in-plane geometries such as photonic crystals or suspended channel waveguides). Both strategies heavily rely on the engineering of the radiation characteristics of QDs by using photonic nanostructures.

\subsection{Extraction to free-space for fiber coupling}

\begin{center}
	\begin{figure}
		\includegraphics[width=\linewidth]{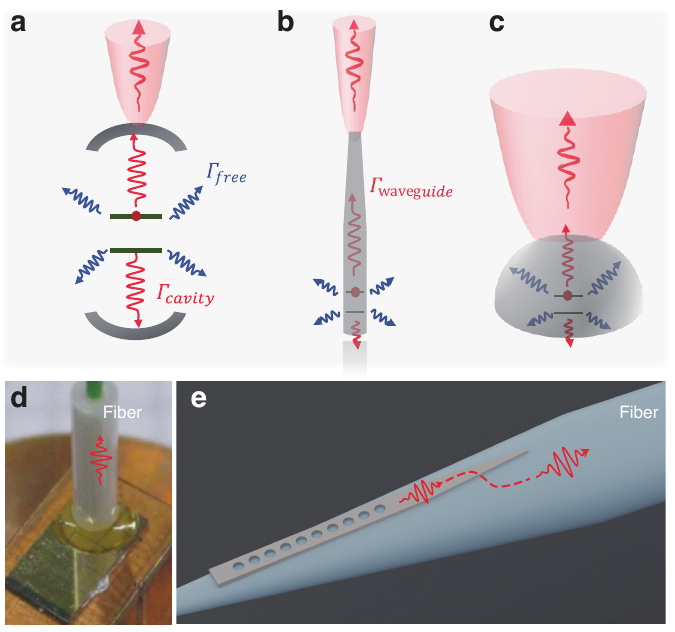}
		\caption{\textbf{Coupling telecom band quantum light sources into single-mode fibers.} (a) Sketches of microcavities used for extracting single photons from semiconductors and coupling into single-mode fiber. The red arrows point out the photon extraction directions,while the blue arrows represent spontaneous emission in other directions; (b) Sketches of photonic waveguides, in which the photons predominantly funnel to the fundamental waveguide mode with high-efficiency light extraction (red arrows). The waveguide end can be carefully tapered to lead the mode to adiabatically expand outside and form a Gaussian-like far-field pattern that can be efficiently collected by a single-mode optical fiber; (c) a typical micro-lens geometry, which can shape the propagation of the emitted photons, effectively guiding the photons towards the top collection optics (red and blue arrows); (d) Photograph of a fiber-coupled QD sample after the epoxide adhesive bonding process\cite{schlehahn2018stand}; (e) Schematic of the integrated photonic crystal nanobeam and the fiber taper for near-field couplings. Reprinted with permission from Ref. [\cite{schlehahn2018stand}].}
		\label{fig:Fig4}
	\end{figure}
\end{center}

The photonic nanostructures for extracting light to the free-space or into optical fibers fall into three categories: optical microcavities, photonic waveguides and lens-based geometry, as schematically shown in Fig.~\ref{fig:Fig4}a-c respectively. An optical microcavity features an advantage of both improving the photon extraction efficiency and accelerating the spontaneous emission rate via the Purcell effect. Microcavities based on pillars\cite{chen2017bright}, photonic crystal defects \cite{kim2016two, birowosuto2012fast,balet2007enhanced} and CBGs\cite{nawrath2022high,xu2022bright,kolatschek2021bright,barbiero2022high,nawrath2023scalable} have been successfully employed to develop bright single-photon sources at both O- and C-bands. In particular, L3 type photonic crystal cavity in O-band exhibits a photon extraction efficiency of (36 $\pm$ 5)$\%$, a high purity of 91.5$\%$ and a high post-selected indistinguishability of 67$\%$\ (raw visibility: 18\%)\cite{kim2016two}, while the O-band (C-band) CBG shows a photon extraction efficiency of 23$\%$\cite{kolatschek2021bright}(16.6$\%$\cite{nawrath2023scalable}), a low $g^{(2)}(0)$ of 0.01\cite{kolatschek2021bright} (0.003\cite{nawrath2023scalable}) with a raw indistinguishability of 19.3$\%$\cite{nawrath2023scalable}. Recently, a simultaneous high-brightness with a fiber-coupled single-photon rate of 13.9 MHz for an excitation rate of 228~MHz has been demonstrated with a CBG while maintaining a low multi-photon contribution of $g^{(2)}(0)$ = 0.005\cite{nawrath2022high}. In a wide-band photonic waveguide\cite{takemoto2007optical,haffouz2020single,takemoto2008telecom,jaffal2019inas,lee2020bright}, spontaneous emission to the radiation modes are greatly suppressed and can only couple to the propagating waveguide mode, improving the photon extraction efficiency with a record value of 27$\%$ for waveguide sources at O-band\cite{lee2020bright}. Without modifying the spontaneous emission dynamics, solid immersion lenses (SILs)\cite{yang2020quantum}, as well as microlens/micromesa-based structures\cite{sartison2018deterministic,musial2021inp}, are another alternative to increase the photon extraction efficiency over a broadband range. Such structures can shape the propagation of the emitted photons, effectively guiding the photons towards the top collection optics. High experimental photon extraction efficiency of 17$\%$ has been reported when GaP SILs\cite{yang2020quantum} or microlens\cite{sartison2018deterministic} are applied to enhance photon extraction of O-band InGaAs/GaAs QDs.

The above has focused on the photonic geometries used to extract QD emission into free-space before coupling into a fiber, and many efforts continue to determine how to best optimize such structure\cite{bremer2022numerical}. Of additional importance are the methods for on-chip fiber coupling, in some cases using free-space micro-optics (i.e., lenses) for mode matching, so that user-friendly fully fiber-coupled sources suitable for alignment-free operation can be realized. One process to realize such functionality is based on active optical alignment and epoxide adhesive bonding at room-temperature\cite{schlehahn2018stand}, as shown in Fig.~\ref{fig:Fig4}d. Using this process, A. Musial et al. deterministically integrated a QD into a micromesa and on-chip coupled to a high numerical aperture (NA) customized single-mode fiber. The realized source can deliver a single-photon emission rate of up to 73 kHz in an application oriented stand-alone device\cite{musial2020plug}.

\subsection{Direct evanescent fiber coupling}

Alternately, single photons emitted by QDs could be directly coupled to optical fibers relies on evanescent wave interactions. When the modes of the two waveguiding structures (one the QD-containing semiconductor waveguide, the other the optical fiber) can be phase-matched, appreciable power transfer - in principle 100\% - can be achieved with sufficient transverse mode overlap and interaction length along the propagation direction\cite{yariv2007photonics}. To realize appreciable transverse mode overlap, tapered optical fibers can be utilized to extract single photons confined in semiconductors into telecom band single-mode fibers. C. M. Lee et al.\cite{lee2019fiber} demonstrate an alignment-free fiber-coupled SPS at telecom wavelength with a brightness of 1.4\% via transferring a PC nanobeam to a tapered optical fiber, as illustrated in Fig.~\ref{fig:Fig4}e. This evanescent coupling scheme can also be implemented in NW waveguides\cite{northeast2021optical}, with a source fiber collection efficiency of 35\% and an overall single photon collection efficiency of 10\%. It is important to note that the direct, on-chip fiber coupling approach usually suffers from much lower additional loss than a typical free space photoluminescence setup, thereby resulting in a similar detected photon flux. Finally, such approaches can be readily applied to QD-containing cavities; the fiber taper can be directly coupled to the cavity, as in Ref. \cite{ates2013improving}, or instead coupled to an on-chip waveguide that is in turn coupled to the cavity, as has been demonstrated in the context of other integrated photonics experiments, where fiber-to-cavity coupling efficiencies as high as 85\% have been shown\cite{groblacher2013highly}.

\begin{table*}[ht!]
	\caption{State of the art of  single-photon sources and entangled photon pair sources based on epitaxial quantum dot}
	\vspace{0.1cm}
	\label{table1}
\renewcommand{\arraystretch}{1.3}
	\centering
	\begin{tabular}{|C{1.3cm}|C{1.6cm}|C{2.2 cm}|C{1.6cm}|C{2.0cm}|C{1.9cm}|C{2.3cm}|C{2.6cm}|C{0.6cm}|}
		\hline
  &\textbf{Wavelength} & \textbf{QDs and growth methods} & \textbf{Photonic structure} & \textbf{Excitation schemes} & \textbf{Photon extraction efficiency} &$g^{(2)}(0)$ & \textbf{Indistinguishability with/without a post-selection} & \textbf{Ref.} \\
		\hline
        \hline
		\multirow{7}{1.5cm}{\\Single-photon sources}  &	O-band & InAs/InP, S-K & Photonic crystals & wetting layer & (36 $\pm$ 5)\% & 0.085 $\pm$ 0.022 & (67.0 $\pm$ 2.0)\%  /          (18.0 $\pm$ 1.0)\% &  \cite{{kim2016two}}\\
\cline{2-9}	
 &	O-band & InAs/InP, S-K & Nanobeam  & p-shell & (27.0 $\pm$ 0.1)\% & 0.077 $\pm$ 0.011 & (91.0 $\pm$ 9.0)\%   /  (20.0 $\pm$ 4.0)\%&  \cite{{lee2020bright}}\\
  \cline{2-9}   
&	O-band & InGaAs/GaAs, SRL & Microlens & wetting-layer & (17.8 $\pm$ 3.8)\% &  0.21 $\pm$ 0.07 & - & \cite{sartison2018deterministic}\\
	\cline{2-9}	
 &	O-band & InGaAs/GaAs, SRL & SIL & above-band & 17\% &  0.049 $\pm$ 0.02 & -& \cite{yang2020quantum} \\
\cline{2-9}
&	O-band & InGaAs/GaAs, SRL & CBG & p-shell &  23\% &  0.01 & - &  \cite{kolatschek2021bright}\\
\cline{2-9} 
 &	C-band & InAs/InP, S-K & CBG & LO-phonon-assisted & (16.6 $\pm$ 2.7)\% & 0.0032 $\pm$ 0.0006 &   (99.8 $^{+0.2}_{-2.6}$ )\% /   (19.3 $\pm$ 2.6)\% &  \cite{{nawrath2023scalable}}\\
 \cline{2-9}

&	C-band & InGaAs/GaAs, MMBL & CBG & p-shell &  17\% &  0.005 & - &  \cite{nawrath2022high}\\
\cline{2-9}

		  &	900 nm & InAs/GaAs, S-K & Open Fabry-Perot cavity  & resonance excitation & (82 $\pm$ 2)\% & 0.0021 & - / 96.7\% & \cite{{tomm2021bright}}\\
		\hline
		&\textbf{Wavelength} & \textbf{QDs and growth methods}& \textbf{Excitation schemes} &\textbf{ Photon pair generation	efficiency} &  \textbf{Peak fidelity} & \textbf{A pair collection probability} & \textbf{Indistinguishability}  & \textbf{Ref.} \\
		\hline
        \hline
		\multirow{2}{1.5cm}{\\Entangled photon pairs}  &	C-band & InAs/GaAs, MMBL & TPE  & (69.9 $\pm$ 3.6)\% & (95.2 $\pm$ 1.1)\% & - & - & \cite{zeuner2021demand}\\
	\cline{2-9}
		  &	780 nm & GaAs/AlGaAs, DE & TPE  & 90\% & (88 $\pm$ 2)\% & (65 $\pm$ 4)\%  & (90.3 $\pm$ 0.3)\% & \cite{liu2019solid}\\
		\hline

	\end{tabular}
\end{table*}

\section{State of the art of epitaxial telecom wavelength quantum light sources and spin-photon interfaces}

In the previous sections, we have extensively discussed techniques used to engineer the extraction efficiency and the in-fiber brightness properties of QDs emitting at telecom wavelength. However, the brightness, single-photon purity, and indistinguishability of telecom band QD SPSs, have only been demonstrated separately in previous experiments. Recent experiments exploiting advanced excitation schemes, as summarized in Table $\rm \uppercase\expandafter{\romannumeral1}$., have demonstrated steady improvements of the device performances, as compared with the QD-based quantum light sources that emitted in the NIR regime.

Epitaxial telecom wavelength quantum light sources have been applied in some proof-of-principle experiments of implementing quantum networks, for example, distributing single photons\cite{Gyger2022Metropolitan} and entangled photons\cite{Xiang2019long,Xiang2020tunable} over deployed telecom fiber that enable wavelength division multiplexing, teleporting with telecom-wavelength polarization-encoded laser qubits\cite{Anderson2020quantum}, random number generation\cite{Gyger2022Metropolitan}, and GHz-clocked quantum relays\cite{PhysRevApplied.13.054052}. These QD-in-photonic structures were further used to demonstrate hundreds of kilometers QKD\cite{Takemoto2015quantum}, overcoming the distance limit of a weak coherent pulse (WCP) source (without decoy states) and a QD-generated single-photon pulse at a wavelength of 900 nm.

There are still a few remaining issues to be solved before obtaining the all-round SPSs at telecom band for multi-photon applications and long-haul entanglement distribution in quantum networks. First, the demonstrated collection efficiency into the first lens is limited by 36\%\cite{kim2016two}, hybrid CBG devices in which a CBG sits on top a thin $\rm SiO_2$ spacer layer with a gold reflector are being exploited for reaching a collection efficiency into the first lens towards 80\%\cite{rickert2019optimized}. Second, most of the reported QD sources at telecom bands featured appreciable blinking and spectral diffusion, which is probably due to the fluctuating charge environment. A p-i-n diode design with ability of tuning the charge states of the QDs and stabilizing the surrounding charged environment is supposed to solve the problem, and there have been significant efforts towards this direction in the community\cite{tomm2021bright,barbiero2022design}. Finally, two photon interference (TPI) with a long-time delay is essential for scalable photonic quantum applications. Telecom single photons with coherence times much longer than the Fourier limit and TPI after 25 km fibre spool \cite{well2022coherent} have been demonstrated on InAs/InP QD platform. Taking a further step, charge environment control using the p-i-n diode\cite{zhai2020low} and fast photon emission from Purcell enhancement\cite{wang2016near} is expected to enable the long-time delay TPI.  For entangled photon pairs at telecom wavelengths, a step forward is fine-structure suppression for S-K type QDs using micromachined piezoelectric actuators\cite{lettner2021strain}, which improving the fidelity of entangled state.

One exciting progress in the community is the very recently demonstrated C-band spin-photon interface based on single InAs/GaAs QD grown on an MMBL\cite{dusanowski2022optical}. The optical spin injection, initialization, read-out, and full coherent control of a confined hole are realized by using picosecond optical pulses. The dephasing time was limited to 240 ps which could be significantly improved in the future for spin-photon and spin-spin entanglement. Integration of QDs into photonic nanostructures can also further significantly enhance the spin-photon coupling and improve the performance of quantum interfaces. 

\section{Quantum frequency conversion of single QDs to telecom wavelengths}

Despite the significant progress that has been made in shifting the emission wavelength of the QD emitters to the telecom band via controlling the epitaxial growth processes, the above-mentioned QDs emitting at telecom bands are generally less mature than their shorter wavelength counterparts, and thus often suffer from high dislocation densities or impurity concentration, resulting in high charge noise, blinking, and therefore low photon indistinguishability. Another route to bridge the frequency gap is QFC, which can convert NIR single photons to the telecom wavelengths while preserving its superior quantum optical properties, been implemented in different quantum systems to realize spin-photon\cite{Tchebotareva2019entangement}/photon-photon entanglement\cite{you2021quantum} or access to quantum memory\cite{arenskotter2022telecom} at telecom range. Compared with other wavelength-tuning approaches used in QD systems such as temperature, electric field or strain, QFC does not influence the emission process itself and therefore can be applied to transfer single photons over a wider spectral range. For quantum internet applications, firstly, QFC can translate the mature NIR single photon to telecom wavelengths with high efficiencies and low noise, allowing high quality SPSs for long-distance QKD. Secondly, QFC can erase the frequency difference and spectral distinguishability among different QDs or with other quantum nodes, which is essential for TPI, a core ingredient in entanglement swapping protocols that, for example, extend the distance of a quantum link.Most importantly, both QD-based quantum devices and QFC modules are increasingly being realized with mature, semiconductor-based planar fabrication technology, making them particularly appealing for on-chip integration.

In this section, we will focus on the near-visible-to-telecom QFC technologies that are suitable for QD quantum light sources, including the second-order nonlinear process based on mature, cm-scale periodically poled lithium niobate (PPLN) waveguides as well as the third-order nonlinear process based on emerging, ultra-compact chip-integrated microresonators ($\rm SiN_x$, AlN, AlGaAs, etc.) which have great potential for implementing QFC on a chip and the possibility of scalability and integration with other chip-integrated photonics.

\subsection{QFC based on cm-scale PPLN waveguides}

\begin{center}
	\begin{figure*}
		\includegraphics[width=0.75\linewidth]{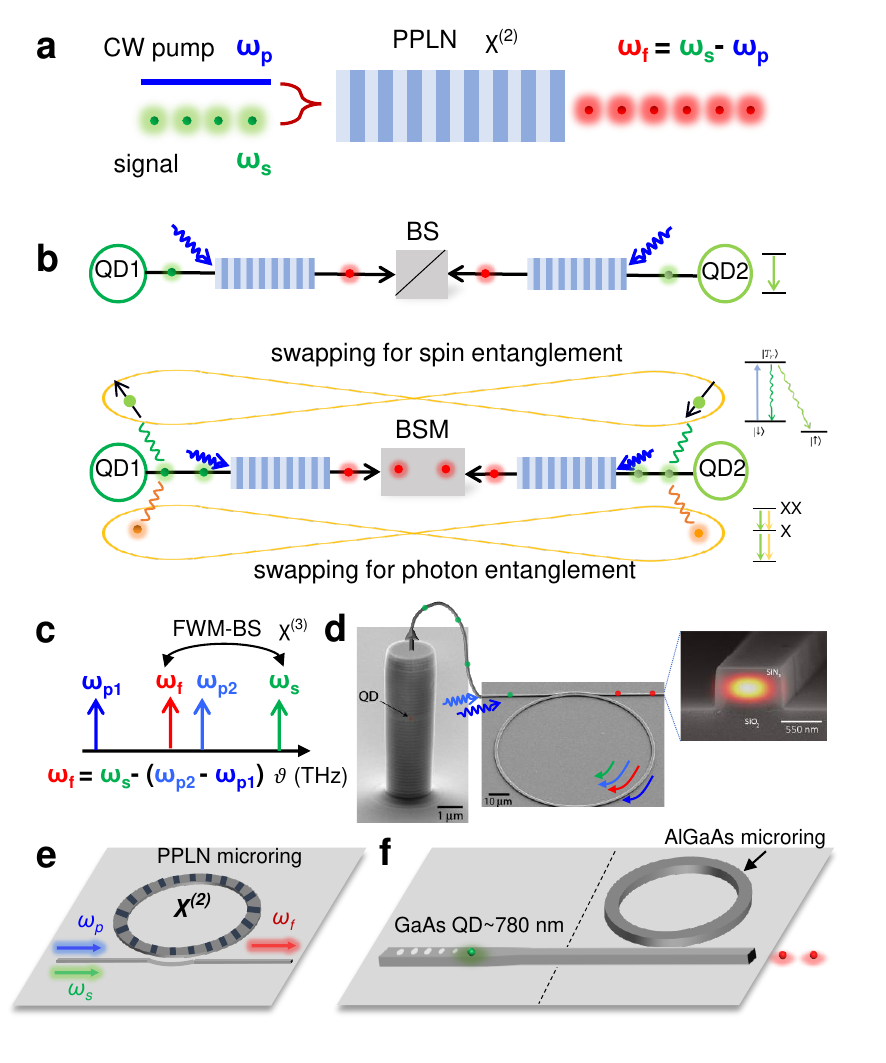}
		\caption{\textbf{Generating telecom band single photons via quantum frequency conversion (QFC).} (a) QFC in periodically poled lithium niobate (PPLN) waveguide through the second-order nonlinear process; (b) The scheme of two-photon interference using QFC telecom single photon (top), as well as the entanglement swapping between the two entangled photon pairs (bottom). (c) Four-wave mixing Bragg scattering (FWM-BS) for visible-telecom QFC; (d) Quantum frequency conversion using the FWM-BS in a hybrid system interfacing $\rm \uppercase\expandafter{\romannumeral3}$-$\rm \uppercase\expandafter{\romannumeral5}$ quantum emitters and nonlinear silicon nitride resonators\cite{singh2019quantum}, together with the SEM image from fabricated $\rm SiN_x$ waveguide with superimposed fundamental TE mode profile at 1550 nm\cite{li2016efficient}; (e) Integrated thin film PPLN circuits for highly-efficient on-chip QFC\cite{chen2021photon}; (f) Integrated GaAs QD based quantum light sources with an AlGaAs microring frequency convertor. Reprinted with permission from Refs. [\cite{li2016efficient,singh2019quantum,chen2021photon}].}
		\label{fig:Fig5}
	\end{figure*}
\end{center}

The QFC enables changing the frequency of the photonic qubits (i.e, the QD SPS, also referred to as the ‘signal’) to the target output frequency (‘idler’) while preserving the quantum information they encode\cite{fejer1994nonlinear}.
PPLN waveguides have served as the most popular choice for QFC. As illustrated in Fig.~\ref{fig:Fig5}a, the QFC is done through the second-order nonlinear process with a $\chi^{(2)}$ nonlinearity in a three-wave mixing process, where a single-photon input is mixed with a strong pump laser beam to produce a single-photon output at either the sum or difference frequency. Such a technique has been widely applied to downshift single photons emitted by QDs to the telecom C-band, where the use of a long wavelength pump near 2000 nm ensures limited noise photon generation. The best reported end-to-end system efficiency is 40.8\% and with a high purity of 97.6\% and a high degree of indistinguishability of 94.8\%\cite{da2022pure}. QFC enabled by the PPLN waveguide has also been used to convert two single photons with an appreciable energy difference from 910 nm to 1560 nm single photons, which erases the which-way frequency information and enabled the realization of spin-photon entanglement\cite{de2012quantum}.

Via the QFC technique, TPIs in the telecom C-band by exploiting QFC on two remote QD emitters were reported with increasingly improved performances\cite{weber2019two,you2021quantum}. A very desirable forward step in QFC is the development of techniques capable of converting one photon of an entangled photon pair source to the telecom bands for entanglement swapping between remote QDs, as illustrated in Fig.~\ref{fig:Fig5}b. The most crucial issue is to preserve the input polarization states. A polarization-insensitive QFC (PIQFC) can be implemented, in which a PPLN waveguide is installed in a Sagnac interferometer\cite{ikuta2018polarization}. Moreover, applying QFC on the photon part of the spin-photon entanglement for TPI interface over the long fiber links could enable the generation of entanglement between remote local solid-state spin qubits, showing prospects for the implementation of quantum memories and multi-photon-entangled cluster states.

\begin{center}
	\begin{figure*}
		\includegraphics[width=0.8\linewidth]{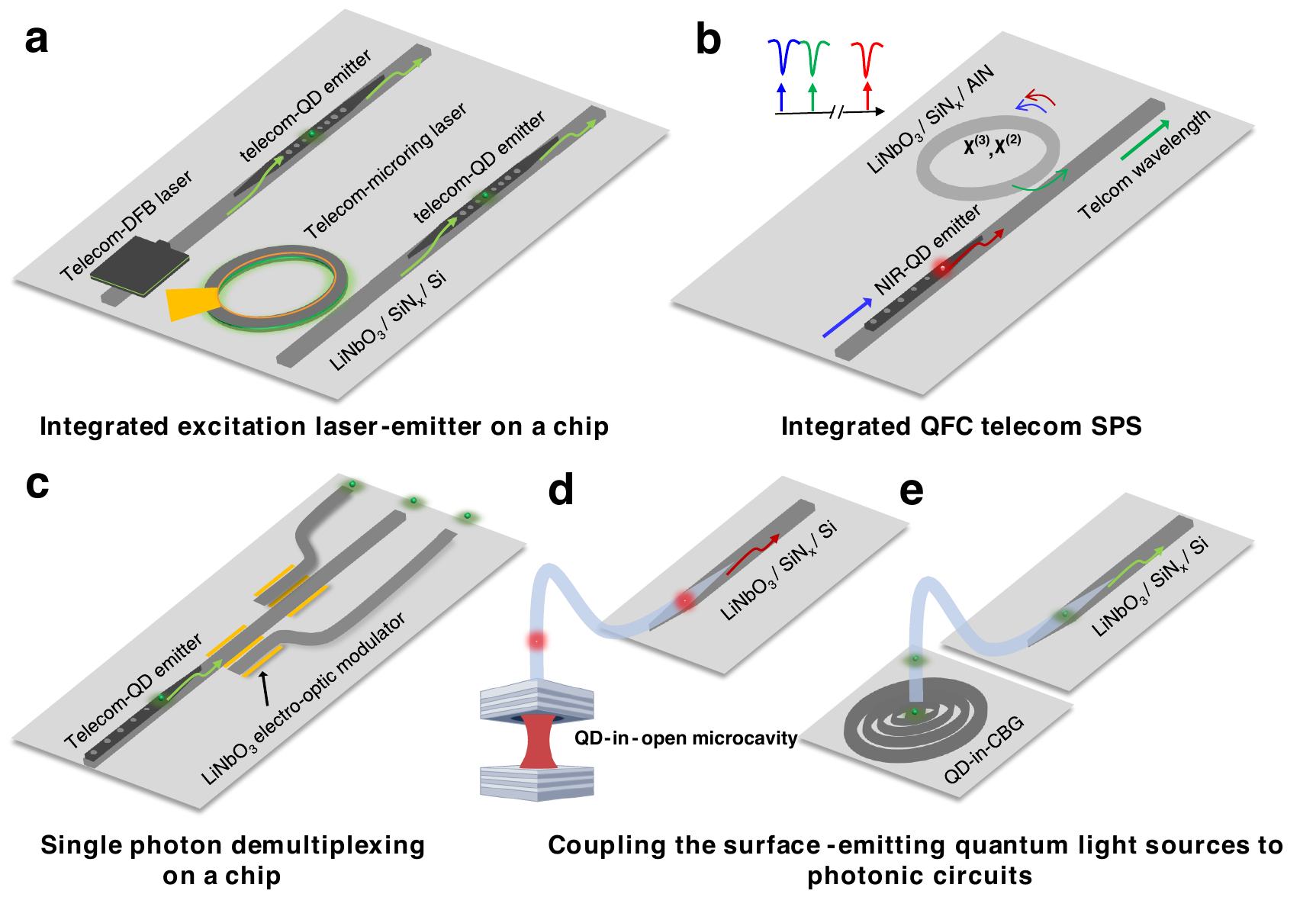}
		\caption{\textbf{Advanced telecom band quantum photonic devices based on hybrid integrations.} (a-c) The sketches of hybrid waveguide-type quantum light sources: telecom band quantum light source with on-chip resonant excitation lasers (a), integrated telecom single photon or entangled photon emitters using on-chip QFC (b), and single photon multiplexing using high performance $\rm LiNbO_3$ electro-optic modulator (c); (d-e) The sketches of hybrid surface emitting quantum light sources via photonic wire bonding technique: integrating with QD in an open-microcavity (d) or a CBG (e).}
		\label{fig:Fig6}
	\end{figure*}
\end{center}

\subsection{QFC based on integrated nanophotonics}

Notably, the strong pump power (on the order of a few hundred mW) and a relatively large size (waveguide length on the order of cm) required by the conventional PPLN can be alleviated by the integrated photonic approach\cite{liu2022emerging,li2016efficient}. In contrast, the large refractive index difference between the nanophotonic waveguide material and the underlying insulator provides tight optical confinement, strong mode overlaps, and new opportunities for dispersion control\cite{moody20222022}. The microring cavity-based structure can further boost the efficiency of second-order or third-nonlinear processes and therefore reduce the required pump power to levels consistent with low power, continuous-wave lasers.

Among the material platforms widely employed in integrated photonics, silicon nitride $\rm SiN_x$\cite{aghaeimeibodi2019silicon} offers broadband optical transparency, relatively high optical nonlinearity, and low linear and nonlinear absorption, which is particularly suitable for QFC experiments. Four-wave mixing Bragg scattering (FWM-BS), a third-order nonlinearity ($\chi^{(3)}$)  process, has been under heavy investigations for low-noise QFC firstly in the optical fibers\cite{mcguinness2010quantum} and then in chip-scale devices\cite{li2016efficient}. In comparison to four-wave mixing processes based on parametric amplification that are often used in classical applications where the generation of added noise photons may not be as important, FWM-BS directly converts signal photons to idler photons, and has no fundamental additive noise floor, thus making it particularly suitable for QFC. As illustrated in Fig.~\ref{fig:Fig5}c, the signal photon will be 'scattered' to idlers (in the frequency domain), with the frequency shift determined by the frequency difference of the two nondegenerate pump lasers. By carefully engineering the phase matching conditions and employing cavity-enhancements, efficient and low-noise intra-band (few nanometers spectral shift) and inter-band (few hundred nanometers spectral shift) QFCs have been demonstrated in $\rm SiN_x$ micro-rings at the single-photon level\cite{li2016efficient}. Such a scheme has been successfully applied to the Fock-states of single photons emitted by the micropillar QD devices for intra-band QFC\cite{singh2019quantum}, as shown in Fig.~\ref{fig:Fig5}d. While in these works the QFC chip and QD chip were separate elements, significant advances in hybrid and heterogeneous integration of QDs (to be discussed in the next section), including with  $\rm SiN_x$ photonics, points to the potential for realizing QD single photon generation and QFC on a single chip in the future. Such integration could significantly improve output photon flux, by removing intermediate chip-to-fiber and fiber-to-chip coupling steps.

The recent technological breakthroughs in integrated photonics could further improve the performances of on-chip QFC devices. For example, the fully-optimized PPLN micro-ring could be used as an ultra-efficient on-chip frequency convertor for single photons\cite{chen2021photon} (Fig.~\ref{fig:Fig5}e). Another emerging opportunity is the recently-developed AlGaAs on insulator platform. The combinations of large $\chi^{(3)}$ nonlinearity\cite{ho1991large}, low propagation loss and the avoidance of two-photon absorptions at C-band wavelengths make it extremely promising for high-performance QFC on-chip. In addition, the material compatibility of AlGaAs with GaAs/AlGaAs droplet-etched QDs\cite{zhai2020low} could enable efficient single-photon generations and QFC on the same chip, as illustrated in Fig.~\ref{fig:Fig5}f. Such a device could avoid the coupling of single photons from sources on one chip to frequency converter on another chip, dramatically improving the overall efficiency of the QFC quantum light sources. To maintain high-conversion efficiencies on-chip, photonic engineering is required to maximize the nonlinearly interacting optical modes at different wavelengths. For the $\chi^{(2)}$ process, e.g., Fig.~\ref{fig:Fig5}e, periodic poling of the sub-$\mu$m domain has recently been demonstrated in TFLN naonophotonic devices \cite{chen2021photon}, dramatically improving the three-wave mixing process for the QFC. In a QFC process based on $\chi^{(3)}$, for instance FWM-BS shown in Fig.~\ref{fig:Fig5}f, the geometry of the waveguide and cladding material can be carefully engineered so that the pumping lasers, signal and idler are simultaneously resonant with different cavity modes associated to the microring resonator, as successfully demonstrated for both short range (a few nm in 980 nm band) and long-range (980 nm to 1550 nm) QFC processes in $\rm SiN_x$ integrated convertors\cite{li2016efficient}.

\section{Hybrid integrated quantum light sources at telecom wavelengths}

Currently, hybrid integrated quantum photonic circuits are an emerging research frontier, offering opportunities of bringing together quantum emitters, quantum memories, coherent linear and nonlinear operations, and single photon detectors on the same chip. The NIR quantum light sources have been benefited from the hybrid integration via pick-and-place technique and wafer-bonding method, as extensively discussed in recent reviews\cite{kim2020hybrid,elshaari2020hybrid}. At telecom bands, pre-characterized single InAs/InP QD SPS have been selectively transferred either on a silicon\cite{kim2017hybrid,aghaeimeibodi2019silicon,ota2021hybrid} or a $\rm LiNbO_3$\cite{aghaeimeibodi2018integration} photonic chips by using a micrometer or submicrometer-sized probe tip\cite{kim2017hybrid} or an adhesive and transparent rubber stamp\cite{ota2021hybrid}. However, the hybrid architecture using wafer bonding\cite{davanco2017heterogeneous} has not yet been reported for quantum light source at telecom wavelengths. One remaining problem for this technique is the random position of the emitters and therefore the actual coupling efficiency and yield remain low. Site-controlled emitters mentioned above and the positioning techniques adapted to telecom bands\cite{xu2022bright} should be able to resolve these issues in the near future.

Proceeding along this path, hybrid integration may bring a new possibility of realizing functional devices involving different wavelengths of QD emitters, full-blown telecom distributed feedback (DFB)/microring laser, a microresonator quantum frequency convertor, and an electro-optic modulator on a chip, as shown in Figs.~\ref{fig:Fig6}a-c. It will be particularly interesting for telecom band quantum light sources because the system can provide high-quality micro-lasers for resonant excitations on the same chip (Fig.~\ref{fig:Fig6}a), integrated telecom single photon or entangle photon emitters using on-chip QFC (Fig.~\ref{fig:Fig6}b), as well as single-photon multiplexing using high performance $\rm LiNbO_3$ electro-optic modulator (Fig.~\ref{fig:Fig6}c).

Besides, photonic wire bonding\cite{billah2018hybrid,Gehring2019lowloss} may offer a solution for efficient coupling from the light sources to photonic circuits. This is achieved by exploiting direct-write two-photon lithography for in situ fabrication of three-dimensional freeform waveguides/couplers between optical chips (Figs.~\ref{fig:Fig6}d-e). This technique may provide a universal integration platform for hybrid photonic multi-chip assemblies that combine known-good devices of different materials to high-performance hybrid multi-chips.

\section{Perspective}

Although tremendous progress has been made towards the development of high-performance telecom wavelength QDs for applications in fiber-based QKD networks and quantum repeaters, there are still important milestone experiments to be pursuited to fully release the potential of telecom QD sources. The near-term goal is the experimental demonstration of beating decoy state protocol by using telecom QD sources in a QKD-based quantum communiciation process. By developing telecom band spin-photon interface with high photon extraction efficiency, spin-spin entanglement over 100 kilometers could be envisioned via entanglement swapping. In the long run, deterministic sources of cluster states with entanglement length up to 50-photons and a generation rate of 2~GHz could enable the realization of high-dimensional cluster states via cluster fusion, which provides the platform for proof-of-principle experiments for high-rate distribution of entanglements between remote nodes.

The realization of these goals will require both the optimization of existing concepts and the development of new techniques beyond those currently available. For epitaxially grown QDs acting as quantum light sources and spin-photon multi-entanglement cluster states, it is important to improve the single-emitter properties, especially in terms of emission linewidth and photon indistinguishability. The latter is still limited to 40-60\% because of enhanced spectral diffusion in O- and C-band QDs due to the detrimental influence of charged defect states in the vicinity of the QDs. Further improvements of the epitaxial layer design and the growth parameter, or even new disruptive growth concepts, are required to overcome these issues and to realize high-performance QDs with Fourier-limited emission linewidths and optimum quantum properties.

Moreover, funneling emission into selected, desirable spatial modes of quantum photonic structures is crucial to maximizing the photon extraction efficiency. For this purpose, deterministic device fabrication technologies for single QD devices, such as various QD positioning and in situ lithography techniques\cite{liu2021nanoscale} used mainly for quantum light sources emitting at $\lambda$ \textless 1000 nm, must be extended to process quantum devices emitting at telecom wavelengths. Careful consideration must be paid to the design and fabrication of the photonic structures so that the geometry and the necessary fabrication process minimally affects the as-grown QD optical properties, such as fabrication induced spectral diffusion which limits the photon indistinguishability\cite{liu2018single}.

For QFC quantum light sources, the development of nonlinear candidates capable of high-quality frequency conversion to telecom wavelengths and their integrations with the existing high-performance quantum light sources can be envisaged. The exceptional nonlinear properties of AlGaAs as well as the attractive bright GaAs/AlGaAs QDs at 780 nm have allowed progress with all of these key components and may offer exciting prospects for high performance QFC quantum light sources at telecom wavelengths on a chip. Low-loss thin film $\rm LiNbO_3$ and $\rm SiN_x$ have also allowed optical nonlinearities at the single photon level, and offered clear advantages with respect to the scalability of the underlying manufacturing processes via the integration with pump lasers and other photonic components. With further technological developments in material growth, photonic design, and integration method, the performances of the telecom band QD devices are steadily improved as quantum light sources and spin-photon interfaces, which plays an increasingly important role in building future solid-state quantum network based on fiber networks.

The cryogenic working temperature ensures the coherent properties of quantum state of the light emitted by QDs. However, for certain applications, e.g., BB84 QKD, room temperature operation is highly preferred for practical reasons while there is no request for the photon coherence and photon indistinguishability. Harnessing exciton localization at defect sites in wide-bandgap semiconductor materials, such as silicon carbide (SiC)\cite{wang2018bright} and gallium nitride (GaN)\cite{meunier2023telecom}, is rapidly emerging as an alternative means to generate single photons at room temperature. Toward a quantum memory, Er$^{3+}$ ions, which shows both long spin and indistinguishable single photon emission\cite{ourari2023indistinguishable}, represent a promising resource for telecom quantum repeater.

\noindent \textbf{Acknowledgements}
We thank the early contributions to this work from Shiwen Xu and Xiaoying Huang. Y. Y. and J. L. acknowledges the National Key R\&D Program of China (2018YFA0306100), the Science and Technology Program of Guangzhou (202103030001), and the National Natural Science Foundation of China (12074442). K. S. acknowledges partial funding support from the NIST-on-a-chip program. S. R. acknowledges financial support from the Federal Ministry of Education and Research (BMBF) via project QR.X (16KISQ014), the German Research Foundation (DFG) via project Re2974-25/1, and the European Union via project SE-QUME (20FUN05) that has received funding from the EMPIR programme co-financed by the Participating States and from the European Union’s Horizon 2020 research and innovation programme. P. M. gratefully acknowledges the funding by the German Federal Ministry of Education and Research (BMBF) via the project QR.X (No.16KISQ013) and the European Union’s Horizon 2020 research and innovation program under Grant Agreement No. 899814 (Qurope). Furthermore, we acknowledge financial support by the EMPIR programme (20FUN05 SEQUME), co-financed by the Participating States and from the European Union’s Horizon 2020 research and innovation programme. C. M. L. and E. W. acknowledge funding support from the National Science Foundation (grant numbers OMA1936314, PHY1915375, ECCS1933546), the Office of Naval Research (ONR) (grant number N000142012551), the Air Force Office of Scientific Research (AFOSR) (grant number 1021098624), the Maryland ARL quantum partnership (MAQP), and the Quantum Leap Challenge Institute for Robust Quantum Simulation.



\end{document}